\documentclass[aps,prd,twocolumns,eqsecnum,preprintnumbers,showpacs,amsmath,amssymb]
{revtex4}

\usepackage{graphicx}
\usepackage{dcolumn}
\usepackage{bm}

\begin{document}

\title{SEVERELY INFRARED SINGULAR GLUON PROPAGATOR IN QCD }

\author{V. Gogohia}
\email[]{gogohia@rmki.kfki.hu}

\affiliation{HAS, CRIP, RMKI, Depart. Theor. Phys., Budapest 114,
P.O.B. 49, H-1525, Hungary}

\date{\today}
\begin{abstract}
We have explicitly shown that the infrared structure of the full
gluon propagator in  QCD is an infinite sum over all severe (i.e.,
more singular than $1/ q^2$) infrared singularities. It reflects
the zero momentum modes enhancement effect in the true QCD vacuum.
Its existence exhibits a characteristic mass (the so-called mass
gap), which is responsible for the scale of nonperturbative
dynamics in the QCD ground state. By an infrared renormalization
of a mass gap only, the deep infrared structure of the full gluon
propagator is saturated by the simplest severe infrared
singularity, the famous $(q^2)^{-2}$. So, there is no smooth in
the infrared limit the full gluon propagator. The main dynamical
source of severe infrared singularities in the gluon propagator is
the two-loop skeleton term of the corresponding equation of
motion, which contains the four-gluon vertices only. Taking into
account the distribution nature of severe infrared singularities,
the gluon confinement criterion is formulated in a manifestly
gauge-invariant way.
\end{abstract}



\maketitle

\section{Introduction}

Quantum Chromodynamics (QCD) \cite{1} is widely accepted as a
realistic, quantum field gauge theory of strong interactions not
only at the fundamental (microscopic) quark-gluon level, but at
the hadronic (macroscopic) level as well. The surprising fact,
however, is that after more than thirty years of QCD, we still
don't know exactly the interaction between quarks and gluons. To
know it means that one knows exactly the full gluon propagator,
the quark-gluon proper vertex and the pure gluon proper vertices.
In the weak coupling limit or in the case of heavy quarks only
this interaction is known. In the first case all the
above-mentioned lower and higher Green's functions (propagators
and vertices, respectively) become effectively free ones
multiplied by the renormalization group corresponding perturbative
(PT) logarithm improvements. In the case of heavy quarks all the
Green's functions can be approximated by their free counterparts
from the very beginning. In general, the Green's functions are
essentially different from their free counterparts (substantially
modified) due to the response of the highly nontrivial large scale
structure of the true QCD vacuum. It is just this response which
is taken into account by the full ("dressed") propagators and
vertices (it can be neglected in the weak coupling limit or for
heavy quarks). That is the main reason why they are still unknown.
In other words, it is not enough to know the Lagrangian of the
theory. In QCD it is also necessary and important to know the true
nonperturbative (NP) structure of its ground state (also there
might be symmetries of the Lagrangian which do not coincide with
symmetries of the vacuum). This knowledge can only come from the
investigation of a general system of the dynamical equations of
motion, the so-called Schwinger-Dyson (SD) system of equations
\cite{1,2,3,4}, to which all the Green's functions should satisfy.

The main purpose of this Letter is to fix the infrared (IR)
structure of the full gluon propagator by analyzing the structure
and some general properties of the SD equation for the full gluon
propagator. This is important and is of broad interest, since it
is closely related to the large scale structure of the true QCD
vacuum as emphasized above.

\section{Gluon propagator}

In order to investigate the problem of the true QCD ground state
structure, let us begin with one of the main objects in the
Yang-Mills (YM) sector. The two-point Green's function, describing
the full gluon propagator, is (Euclidean signature here and
everywhere below)

\begin{equation}
D_{\mu\nu}(q) = i \left\{ T_{\mu\nu}(q)d(q^2, \xi) + \xi
L_{\mu\nu}(q) \right\} {1 \over q^2 },
\end{equation}
where $\xi$  is the gauge fixing parameter ($\xi = 0$ - Landau
gauge and  $\xi = 1$ - Feynman gauge) and
$T_{\mu\nu}(q)=\delta_{\mu\nu}-q_{\mu} q_{\nu} / q^2 =
\delta_{\mu\nu } - L_{\mu\nu}(q)$. Evidently, $T_{\mu\nu}(q)$ is
the transverse (physical) component of the full gluon propagator,
while $L_{\mu\nu}(q)$ is its longitudinal (unphysical) one. The
free gluon propagator is obtained by setting simply the full gluon
form factor $d(q^2, \xi)=1$ in Eq. (2.1), i.e.,

\begin{equation}
D^0_{\mu\nu}(q) = i \left\{ T_{\mu\nu}(q) + \xi
L_{\mu\nu}(q) \right\} {1 \over q^2 }.
\end{equation}

 The solutions of the SD equation for the full gluon propagator (2.1)
are supposed to reflect the complexity of the quantum structure of
the QCD ground state. Just this determines one of the central
roles of the full gluon propagator in the SD system of equations
\cite{5}. The SD equation for the full gluon propagator (see Eq.
(3.1) below) is a highly nonlinear system of four-dimensional
integrals, containing many different, unknown in general,
propagators and vertices, which, in their turn, satisfy too
complicated integral equations, containing different scattering
amplitudes and kernels, so there is no hope for exact solution(s).
However, in any case the solutions of this equation can be
distinguished from each other by their behavior in the IR limit,
describing thus many (several) different types of quantum
excitations and fluctuations of gluon field configurations in the
QCD vacuum. Evidently, not all of them can reflect the real
structure of the QCD vacuum, for example the gauge artifact
solutions (see below). The ultraviolet (UV) limit of these
solutions is uniquely determined by asymptotic freedom (AF)
\cite{6}.

The deep IR asymptotics of the full gluon propagator can be
generally classified into the two different types: singular, which
means that the zero momentum modes enhancement (ZMME) effect takes
place in the NP QCD vacuum, or smooth, which means that the full
gluon propagator is IR finite or even is IR vanishing. Formally,
the full gluon propagator (2.1) has an exact power-type IR
singularity, $1/q^2$, which is due to its longitudinal component.
This is the IR singularity of the free gluon propagator, see Eq.
(2.2). So by the ZMME effect we mean, in general, the IR
singularities, which are more severe than $1/q^2$. Evidently, the
singular asymptotics is possible at any value of the gauge fixing
parameter. At the same time, the smooth behavior of the full gluon
propagator (2.1) in the IR becomes formally possible either  by
choosing the Landau gauge $\xi = 0$ from the very beginning, or by
removing the longitudinal (unphysical) component of the full gluon
propagator with the help of ghost degrees of freedom \cite{1,7,8}
(for more detail discussion see below).

 However, any deviation in the behavior of the full gluon propagator
from the free one in the IR domain automatically assumes its
dependence on a scale parameter (at least one) different, in
general, from the QCD asymptotic scale parameter $\Lambda_{QCD}$.
It can be considered as responsible for the NP dynamics (in the IR
region) in the QCD vacuum. If QCD itself is a confining theory,
then such a characteristic scale is very likely to exist. This is
very similar to AF, which requires the above-mentioned asymptotic
scale parameter $\Lambda_{QCD}$ associated with the nontrivial PT
dynamics in the UV region (AF, scale violation, determining thus
the deviation in the behavior of the full gluon propagator from
the free one in the UV domain). In this connection it is worth
emphasizing that, being numerically a few hundred $MeV$ only, it
cannot survive in the UV limit. This means that none of the finite
scale parameters, in particular $\Lambda_{QCD}$, can be determined
by PT QCD. It should come from the IR region, so it is NP by
origin. How to establish a possible relation between these two
independent scale parameters was shown in our paper \cite{9}.
Despite the fact that the PT vacuum cannot be the true QCD ground
state \cite{10}, nevertheless, the existence of such kind of a
relation is a manifestation that "the problems encountered in
perturbation theory are not mere mathematical artifacts but rather
signify deep properties of the full theory" \cite{11}.

The message that we have tried to convey is that precisely AF
clearly indicates the existence of the NP phase with its own
characteristic scale parameter in the full QCD.

\section{Gluon SD equation}

The general structure of the SD equation for the full gluon
propagator can be written down symbolically as follows (for our
purposes it is more convenient to consider the SD equation for the
full gluon propagator and not for its inverse):

\begin{equation}
D(q) = D^0(q) - D^0(q)T_{gh}(q) D(q) - D^0(q)T_q(q)D(q) +
D^0(q)T_g[D](q)D(q).
\end{equation}
Here and in some places below, we omit the dependence on the Dirac
indices, for simplicity. $T_{gh}(q)$ and $T_q(q)$ describe the
ghost and quark skeleton loop contributions into the gluon
propagator. They do not contain the full gluon propagators by
themselves. A pure gluon contribution $T_g[D](q)$ is a sum of four
pure gluon skeleton loops, and consequently they explicitly
contain the full gluon propagators. Precisely this makes the gluon
SD equation highly nonlinear (NL), and this is one of the reasons
why it cannot be solved exactly. However, its linear part, which
contains only ghost and quark skeleton loops, can be summed up, so
Eq. (3.1) becomes

\begin{equation}
D(q) = \tilde{D}^0(q) + \tilde{D}^0(q)T_g[D](q)D(q) = \tilde{D}^0(q) + D^{NL}(q),
\end{equation}
with $\tilde{D}^0(q)$ being a modified free gluon propagator as
follows:

\begin{equation}
\tilde{D}^0(q) = { D^0(q)  \over 1 + [T_{gh}(q) + T_q(q)]D^0(q)},
\end{equation}
where

\begin{equation}
T_{gh}(q) =  g^2 \int {i d^4 k \over (2 \pi)^4} k_{\nu} G(k)
G(k-q)G_{\mu}(k-q, q),
\end{equation}

\begin{equation}
T_q(q) = - g^2 \int {i d^4 p \over (2 \pi)^4} Tr [\gamma_{\nu}
S(p-q) \Gamma_{\mu}(p-q, q)S(p)].
\end{equation}
In general, these quantities can be decomposed as follows:

\begin{equation}
T_{gh}(q) \equiv T^{gh}_{\mu\nu}(q) = \delta_{\mu\nu} q^2
T_{gh}^{(1)}(q^2) + q_{\mu} q_{\nu} T_{gh}^{(2)}(q^2),
\end{equation}

\begin{equation}
T_q(q) \equiv T^q_{\mu\nu}(q) = \delta_{\mu\nu} q^2 T_q^{(1)}(q^2) +
q_{\mu} q_{\nu} T_q^{(2)}(q^2),
\end{equation}
where all invariant functions $T_{gh}^{(n)}(q^2)$ and
$T_q^{(n)}(q^2)$ at $n=1,2$ are dimensionless with a regular
behavior at zero (they include the dependence on the coupling
constant squared $g^2$). In this connection a few remarks are in
order. Due to the definition $q_{\mu} q_{\nu} =q^2 L_{\mu\nu}$,
instead of the independent structures $\delta_{\mu\nu}$ and
$q_{\mu} q_{\nu}$ in Eqs. (3.6) and (3.7), one can use
$T_{\mu\nu}$ and $L_{\mu\nu}$ as independent structures with their
own invariant functions. For simplicity we assume here and
everywhere below that all integrals are finite, and consequently
all invariant functions are also finite at zero. Anyway, how to
render them finite is well known procedure (see, for example Refs.
\cite{1,7,8,12}).

From a technical point of view it is convenient to use the free
gluon propagator (2.2) in the Feynman gauge ($\xi=1$), i.e,
$D^0_{\mu\nu}(q) = \delta_{\mu\nu} (i /q^2)$. Then from Eq. (3.3)
it follows

\begin{equation}
\tilde{D}^0(q) =  D^0(q) A(q^2),
\end{equation}
where $A(q^2)= 1 / (1 + T(q^2))$, and $T(q^2)$ is regular at zero.
Obviously, it is a combination of the previous ghost
$T_{gh}^{(n)}(q^2)$ and quark  $T_q^{(n)}(q^2)$ at $n=1,2$
invariant dimensionless functions (it includes the dependence on
the coupling constant squared again and the gauge fixing parameter
as well in the general case (i.e., when $D^0(q)$ is given by Eq.
(2.2)). Since $A(q^2)$ is finite at zero, the IR singularity of
the linear part of the full gluon propagator is completely
determined by the power-type exact IR singularity of the free
gluon propagator, as it follows from Eq. (3.8), i.e.,
$\tilde{D}^0(q) = A(0) D^0(q), \quad q^2 \rightarrow 0$. We are
especially interested in the structure of the full gluon
propagator in the IR region, so the exact result (3.8) will be
used as an input in the direct iteration solution of the gluon SD
equation (3.2). Evidently, this form of the gluon SD equation
makes it possible to take into account automatically ghost and
quark degrees of freedom in all orders of linear PT. On the other
hand, it emphasizes the important role of the pure gluon
contribution (i.e., YM one), which forms its NL part.

Let us present now explicitly the NL pure gluon part, which was
symbolically denoted as $T_g[D](q)$ in the gluon SD Eq. (3.2). As
mentioned above, it is a sum of four terms, namely

\begin{equation}
T_g[D](q)  = {1 \over 2} T_t + {1 \over 2} T_1(q) + {1 \over 2}
T_2(q) + {1 \over 6} T_2'(q),
\end{equation}
where the corresponding quantities are given explicitly below

\begin{equation}
T_t =  g^2 \int {i d^4 q_1 \over (2 \pi)^4} T^0_4 D(q_1),
\end{equation}

\begin{equation}
T_1(q) =  g^2 \int {i d^4 q_1 \over (2 \pi)^4} T^0_3 (q, -q_1,
q_1-q) T_3 (-q, q_1, q -q_1) D(q_1) D(q -q_1),
\end{equation}

\begin{equation}
T_2(q) =  g^4 \int {i d^4 q_1 \over (2 \pi)^4} \int {i d^n q_2
\over (2 \pi)^4} T^0_4 T_3 (-q_2, q_3, q_2 -q_3) T_3(-q, q_1,
q_3-q_2) D(q_1) D(-q_2)D(q_3) D(q_3 -q_2),
\end{equation}

\begin{equation}
T_2'(q) =  g^4 \int {i d^4 q_1 \over (2 \pi)^4} \int {i d^4 q_2
\over (2 \pi)^4} T^0_4 T_4 (-q, q_1, -q_2, q_3) D(q_1)
D(-q_2)D(q_3).
\end{equation}
In the last two equations $q-q_1 +q_2-q_3=0$ is assumed as usual.
The $T_t$ term, which is given in Eq. (3.10), is the so-called
tadpole term contribution into the gluon propagator (gluon
self-energy). The $T_1(q)$ term describes the one-loop skeleton
contribution, depending on the three-gluon vertices only. The
$T_2(q)$ term describes the two-loop skeleton contribution,
depending on the three- and four-gluon vertices, while the
$T_2'(q)$ term describes the two-loop skeleton contribution,
depending on the four-gluon vertices only.

The formal iteration solution of Eq. (3.2) looks like

\begin{equation}
D(q) = D^{(0)}(q)  + \sum_{k=1}^{\infty}D^{(k)}(q) = D^{(0)}(q) +
\sum_{k=1}^{\infty} \Bigl[ D^{(0)}(q) T_g \Bigl[
\sum_{m=0}^{k-1}D^{(m)} \Bigr] (q) \Bigl( \sum_{m=0}^{k-1}D^{(m)}
(q)\Bigr) - \sum_{m=1}^{k-1}D^{(m)}(q) \Bigr],
\end{equation}
where, for example explicitly the first four terms are:

\begin{eqnarray}
D^{(0)}(q) &=& \tilde{D}^0(q), \nonumber\\
D^{(1)}(q) &=& \tilde{D}^0(q)T_g[\tilde{D}^0] (q) \tilde{D}^0(q), \nonumber\\
D^{(2)}(q) &=& \tilde{D}^0(q)T_g[\tilde{D}^0 +  D^{(1)}] (q)
(\tilde{D}^0(q) + D^{(1)}(q)) - D^{(1)}(q), \nonumber\\
D^{(3)}(q) &=& \tilde{D}^0(q)T_g[\tilde{D}^0 + D^{(1)} + D^{(2)}](q)
(\tilde{D}^0(q) + D^{(1)}(q) + D^{(2)}(q))
 - D^{(1)}(q) - D^{(2)}(q),
\end{eqnarray}
and so on. It is worth mentioning that the order of iteration does
not coincide with the order of PT in the coupling constant
squared. For example, any iteration (even zero) in Eq. (3.14)
contains ghost and quark degrees of freedom in all orders of PT,
as underlined above. In other words, the iteration solution (3.14)
is a general one, since the skeleton loop contributions (skeleton
diagrams) are to be iterated (the so-called general iteration
solution). In principle, it should be distinguished from the pure
PT iteration solution, i.e., from the expansion in powers of the
coupling constant squared. In this case a pure PT diagrams (with
free propagators and point-like vertices) are to be iterated.

\section{The explicit functional estimate}

Let us now establish a type of a possible functional dependence of
the full gluon propagator in the IR region. For this purpose it is
convenient to start with the gluon SD equation (3.2). Up to the
first iteration it becomes

\begin{equation}
D(q) = \tilde{D}^0(q) + \tilde{D}^0(q)T_g[D](q)D(q) =
        \tilde{D}^0(q) +
        \tilde{D}^0(q)T_g[\tilde{D}^0](q)\tilde{D}^0(q)+ .....,
\end{equation}
where we will use Eq. (3.8) for the modified free gluon propagator
in the Feynman gauge in what follows.

 In the first approximation of Eq. (3.13), i.e.,
at the order $g^4$ we put $T_4=T_4^0$, and it becomes

\begin{equation}
T'_2(q) \equiv T'^{(2)}_{\nu_1\mu_1}(q) = g^4 \int {i d^4 q_1
\over (2 \pi)^4} \int {i d^4 q_2 \over (2 \pi)^4}
T^0_{\nu_1\rho\lambda\sigma} T^0_{\mu_1\rho_1\lambda_1\sigma_1}
\tilde{D}^0_{\rho\rho_1}(q_1) \tilde{D}^0_{\lambda\lambda_1}
(-q_2) \tilde{D}^0_{\sigma\sigma_1} (q-q_1+q_2),
\end{equation}
where it is assumed that the summation over color group factors
has been already done and is included into the coupling constant,
since these numbers are not important. The summation over Dirac
indices then yields

\begin{equation}
T'^{(2)}_{\nu_1\mu_1}(q) = - i \delta_{\nu_1\mu_1} g^4 \int
id^4q_1 \int id^4q_2 { A(q_1^2) A(q_2^2) A((q-q_1+q_2)^2)  \over
q_1^2 q_2^2 (q-q_1+q_2)^2 } = - i \delta_{\nu_1\mu_1} g^4
F'_2(q^2),
\end{equation}
where we introduce

\begin{equation}
F'_2(q^2) = \int id^4q_1 \int id^4q_2 { A(q_1^2) A(q_2^2)
A((q-q_1+q_2)^2) \over q_1^2 q_2^2 (q-q_1+q_2)^2 }.
\end{equation}
 In order to introduce a mass gap, which determines the deviation
of the full gluon propagator from the free one in the IR region,
at the level of the separate diagram (contribution), let us
present the last integral as a sum of four terms, namely

\begin{equation}
F'_2(q^2) = \sum_{n=1}^{n=4} F'^{(n)}_2(q^2),
\end{equation}
where

\begin{equation}
F'^{(1)}_2(q^2) = \int_0^{\Delta^2} id^4q_1 \int_0^{\Delta^2}
id^4q_2 {  A(q_1^2) A(q_2^2) A((q-q_1+q_2)^2) \over q_1^2 q_2^2
(q-q_1+q_2)^2 },
\end{equation}

\begin{equation}
F'^{(2)}_2(q^2) = \int_{\Delta^2}^{\infty} id^4q_1
\int_0^{\Delta^2} id^4q_2 {  A(q_1^2) A(q_2^2) A((q-q_1+q_2)^2)
\over q_1^2 q_2^2 (q-q_1+q_2)^2 },
\end{equation}

\begin{equation}
F'^{(3)}_3(q^2) = \int_0^{\Delta^2} id^4q_1
\int_{\Delta^2}^{\infty} id^4q_2 {  A(q_1^2) A(q_2^2)
A((q-q_1+q_2)^2) \over q_1^2 q_2^2 (q-q_1+q_2)^2 },
\end{equation}

\begin{equation}
F'^{(4)}_2(q^2) = \int_{\Delta^2}^{\infty} id^4q_1
\int_{\Delta^2}^{\infty} id^4q_2 {  A(q_1^2) A(q_2^2)
A((q-q_1+q_2)^2) \over q_1^2 q_2^2 (q-q_1+q_2)^2 },
\end{equation}
where not loosing generality we introduced the common mass gap
squared $\Delta^2$ for both loop variables $q_1^2$ and $q_2^2$.
The integration over angular variables is assumed.

 We are especially interested in the region of
all the small gluon momenta involved $q \approx q_1 \approx q_2
\approx 0$. However, in Eq. (4.6) we can formally consider the
variables $q_1$ and $q_2$ as much smaller than the small gluon
momentum $q$, i.e., to approximate $q_1 \approx \delta_1q, \ q_2
\approx \delta_2q$, so that $q-q_1+q_2 \approx q(1+ \delta)$,
where $\delta = \delta_2 - \delta_1$. To leading order in
$\delta$, one obtains

\begin{equation}
F'^{(1)}_2(q^2) = - {A(q^2) \over q^2} \int_0^{\Delta^2} dq^2_1
\int_0^{\Delta^2} dq^2_2 A(q_1^2) A(q_2^2),
\end{equation}
where all the finite numbers after the trivial integration over
angular variables will be included into the numerical factors
below, for simplicity. Since $q^2$ is small, we can replace the
dimensionless function $A(q^2)$ by its Taylor expansion as
follows: $A(q^2) = A(0) + a_1 (q^2 / \Delta^2) + O(q^4)$.
Introducing further dimensionless variables $q_1^2 = x_1 \Delta^2$
and $q_2^2 = x_2 \Delta^2$, one finally obtains

\begin{equation}
F'^{(1)}_2(q^2) = - {\Delta^4 \over q^2} c_1  - \Delta^2 c'_1  +
O(q^2),
\end{equation}
where

\begin{equation}
c_1 = A(0) \int_0^1 dx_1 A(x_1) \int_0^1 dx_2 A(x_2),
\end{equation}
and $c'_1 = a_1 (c_1 /A(0))$.

In Eq. (4.7) it makes sense to approximate $q_2 \approx
\delta_3q_1, \ q \approx \delta_4q_1$, so that $q-q_1+q_2 \approx
q_1(1+ \tilde{\delta})$, where $\tilde{\delta} = \delta_4 -
\delta_3$. To leading order in $\tilde{\delta}$ and omitting some
algebra, one finally obtains

\begin{equation}
F'^{(2)}_2(q^2) = - \Delta^2 c_2(\nu) + O(q^2),
\end{equation}
where

\begin{equation}
c_2(\nu) = \int_1^{\nu} {dx_1 \over x_1} A^2(x_1) \int_0^1 dx_2
A(x_2),
\end{equation}
and $\nu$ is the dimensionless auxiliary UV cut-off.

In Eq. (4.8) it makes sense to approximate $q_1 \approx
\delta_5q_2, \ q \approx \delta_6q_2$, so that $q-q_1+q_2 \approx
q_2(1+ \bar \delta)$, where $\bar \delta = \delta_5 + \delta_6$.
To leading order in $\bar \delta$ and similar to the previous
case, one obtains

\begin{equation}
F'^{(3)}_2(q^2) = - \Delta^2 c_3(\nu) + O(q^2),
\end{equation}
where

\begin{equation}
c_3(\nu) = \int_1^{\nu} {dx_2 \over x_2} A^2(x_2) \int_0^1 dx_1
A(x_1).
\end{equation}
The last term (4.9) is left unchanged, since all loop variables
are big. Conventionally, we will call it as the PT part of the
contribution (diagram), i.e., denoting $F'^{(4)}_2(q^2)$ as
$T'^{PT}_2(q^2)$. Since $A(x)$ is regular at zero, the both
integrals in Eqs. (4.14) and (4.16) are logarithmic divergent.

 Summing up all terms, one obtains

\begin{equation}
T'_2(q) \equiv T'^{(2)}_{\nu_1\mu_1}(q) = i \delta_{\nu_1\mu_1}
\Bigr[ {\Delta^4 \over q^2} c_1 + \Delta^2( c'_1 + c_2(\nu) + c_3
(\nu)) \Bigl] g^4 + O(q^2).
\end{equation}
The term $T'^{PT}_2(q^2)$ is hidden in terms $O(q^2)$. Here the
characteristic mass scale parameter $\Delta^2$ is responsible for
the nontrivial dynamics in the IR domain. Let us also emphasize
that the limit $\nu \rightarrow \infty$ should be taken at the
final stage.

 In the same way can be decomposed the simplest
one-loop contribution of the order $g^2$ into the quark
self-energy provided by a three-gluon vertices insertion, see Eq.
(3.11). Omitting all the algebra, it is instructive to present it
in the similar to Eq. (4.17) form, namely

\begin{equation}
T_1(q) \equiv  T^{(1)}_{\nu_1\mu_1}(q) = - i \delta_{\nu_1\mu_1}
\Delta^2 c_4 g^2 + O(q^2),
\end{equation}
where $c_4$ is some finite number. The contribution of the
$T_2(q)$ term given in Eq. (3.12) can be given by the estimate
similar to the estimate (4.18) with different finite coefficients,
of course. Evidently, instead of $g^2$ it should be multiplied by
$g^4$. Let us note here that in dimensional regularization the
constant tadpole term (3.10) in the pure PT iteration solution
(i.e., at the order $g^2$, which means $D =D^0 + ...$) of the
gluon SD equation (3.1) can be generally discarded \cite{17}.
Thus, it itself is not important at all.

Two principal distinction of the estimate (4.17) from the estimate
(4.18) should be underlined. First, in the latter estimate there
is only constant contribution, i.e., there is no singular with
respect to $q^2$ term. Secondly, this contribution does not depend
on the UV cut-off. This means that in the final limit when the UV
cut-off will go to infinity it will be suppressed. This
observation underlines the role of the four-gluon interactions in
the IR structure of the true QCD vacuum (see below). The
three-gluon proper vertex vanishes when all the gluon independent
momenta involved go to zero, i.e., $T_3(0,0) \rightarrow
T_3^0(0,0) =0$. At the same time the four-gluon proper vertex
survives in this limit, i.e., $T_4(0,0,0) \rightarrow T_4^0(0,0,0)
\neq 0$. This is the main dynamical reason in the above-mentioned
distinction between the functional estimates (4.17) and (4.18).
Evidently, such kind of the auxiliary procedure, described in this
Sec., is appropriate only for the establishing the most singular
terms in the deep IR asymptotics of the full gluon propagator.

\section{Severe IR structure of the QCD vacuum }

At the NL two-loop level, i.e., at the order $g^4$, there is a
number of the additional diagrams, which, however, contain the
three-gluon vertices (plus the two-tadpole diagram) along with the
four-gluon ones. As mentioned above, their contributions into the
deep IR structure of the gluon propagator are given by the
estimates similar to the estimate (4.18) with different
coefficients. Summing up all these estimates, and omitting some
really tedious algebra, the full gluon propagator up to the first
iteration (including all the terms of the order $g^2$ and $g^4$)
can be written as

\begin{equation}
D_{\mu\nu}(q) = i \delta_{\mu\nu} \Bigl[ {\Delta^2 \over (q^2)^2}
a_1 + {\Delta^4 \over (q^2)^3} a_2 + ... \Bigr] +
D^{PT}_{\mu\nu}(q)= D^{INP}_{\mu\nu}(q)+ D^{PT}_{\mu\nu}(q),
\end{equation}
where $a_1, \ a_2$ are, in general, the short-hand notations for a
sums of the different coefficients, which include the coupling
constant squared in the corresponding powers. Moreover, some of
these coefficients contain the divergent integrals (see, for
example Eqs. (4.14) and (4.16)). Here $D^{PT}_{\mu\nu}(q)$ denotes
the contribution from the PT part of the full gluon propagator,
since it is of the order $O(q^{-2})$ as $q^2 \rightarrow 0$. The
superscript "INP" stands for the intrinsically NP part of the full
gluon propagator (for the exact definition see below). Due to the
above-emphasized distinction between the behavior of the tree- and
four-gluon vertices in the deep IR domain, the coefficients $a_1,
\ a_2$ are, in general, not zero. In other words, there is no way
to cancel $D^{INP}_{\mu\nu}(q)$ by performing the functional
estimate at every order of the QCD coupling constant squared. In
the deep IR region the quark and ghost degrees of freedom (the
$A(q^2)$ function) are taken into account in all orders of linear
PT numerically, i.e., they are simply numbers. As functions they
can contribute into the PT part only of the full gluon propagator.
So, in the first approximation the gluon propagates like Eq. (5.1)
and not like the modified free one (3.3), though we just started
from it.

 The true QCD vacuum is really beset with severe (i.e., more singular
than $1/q^2$ as $q^2 \rightarrow 0$) IR singularities if standard
PT is applied. Moreover, each severe IR singularity is to be
accompanied by the corresponding powers of the mass gap,
responsible for the NP dynamics in the IR region. In more
complicated cases of the multi-loop diagrams (i.e., the next
iterations in Eq. (4.1)) more severe IR divergences will appear.
The coefficients at each severe IR singularity become by
themselves an infinite series in the coupling constant squared,
and the coefficients of these expansions may depend on the gauge
fixing parameter as well \cite{7}. These coefficients include
numerically the information about quark and ghost degrees of
freedom in all orders of linear PT, as underlined above.

It is worth emphasizing, however, that the ZMME effect in the QCD
vacuum, which is explicitly shown in Eq. (5.1) in the Feynman
gauge, can be demonstrated in any covariant gauge, for example in
the Landau one $\xi=0$. In other words, this effect itself is
gauge-invariant, though the finite sum of all the relevant
diagrams in the deep IR region at the same order of the coupling
constant squared may be not. Let us also remind that this effect
is not something new. It has been well known for a long time from
the very beginning of QCD, and it was the basis for the proposed
then IR slavery (IRS) mechanism of quark confinement
\cite{1,13,14,15,16,17,18}. Just this violent IR behavior makes
QCD as a whole an IR unstable theory, and therefore it has no IR
stable fixed point, indeed \cite{1,13}.

The existence of a severe IR singularities automatically requires
an introduction of a mass gap, responsible for the nontrivial
dynamics in the IR region. This is important, since there is none
explicitly present in the QCD Lagrangian (the current quark mass
cannot be considered as a mass gap, since it is not
renormalization group invariant). It precisely determines to what
extent the full gluon propagator effectively changes its behavior
from the behavior of the free one in the IR domain. The phenomenon
of ''dimensional transmutation'' \cite{1} only supports our
general conclusion that QCD may exhibit a mass, determining the
characteristic scale of the NP dynamics in its ground state. Of
course, such gluon field configurations, which are to be described
by severely IR structure of the full gluon propagator, can be only
of dynamical origin. The only dynamical mechanism in QCD which can
produce such configurations in the vacuum, is the self-interaction
of massless gluons -- the main dynamical NL effect in QCD. Hence,
the above-mentioned mass gap appears on dynamical ground. Let us
remind that precisely this self-interaction in the UV limit leads
to AF.

We have explicitly shown that the low-frequency components of the
virtual fields in the true vacuum should have larger amplitudes
than those of a PT ("bare") vacuum \cite{19}, indeed. "But it is
to just this violent IR behavior that we must look for the key to
the low energy and large distance hadron phenomena. In particular,
the absence of quarks and other colored objects can only be
understood in terms of the IR divergences in the self-energy of a
color bearing objects" \cite{14}. So, let us introduce the
following definitions:

(i). The power-type IR singularity which is more severe than the
exact power-type IR singularity of the free gluon propagator will
be called a severe (or equivalently NP IR) singularity. In other
words, the NP IR singularity is more severe than $1/ q^2$ at $q^2
\rightarrow 0$.

(ii). At the same time, the IR singularity which is as much
singular as the exact power-type IR singularity of the free gluon
propagator, i.e., as much singular as $1/ q^2$ at $q^2 \rightarrow
0$, will be called PT IR singularity.

It makes worth emphasizing in advance that the decomposition of
the full gluon propagator into the INP and PT parts (5.1) can be
made exact \cite{20,21}. From the distribution theory point of
view the NP IR singularities defined above present a rather broad
and important class of functions with algebraic singularities
\cite{22}. This explicit derivation shows how precisely the NP IR
singularities, accompanied with a mass gap in the corresponding
powers, may appear in the vacuum of QCD. Thus, the NP IR
singularities should be summarized (accumulated) into the full
gluon propagator and effectively correctly described by its
structure in the deep IR domain, presented by its INP part. The
second step is, of course, to assign a mathematical meaning to the
integrals, where such kind of the NP IR singularities will
explicitly appear, i.e., to define them correctly in the IR region
(see the next Subsec.).

{\bf One important thing should be made perfectly clear. In the
exact calculation of a separate diagram the dependence on the
characteristic masses (determining the deviation of the full gluon
propagator from the free one in the IR and UV regions) is hidden.
In other words, these masses cannot be "seen" by the calculation
of the finite number of diagrams, which may be not even
gauge-invariant. An infinite number of the corresponding diagrams
should be summed up in order to trace such NP masses (i.e., to go
beyond PT). The final result of such summation should, in
principle, be gauge-invariant. In the weak coupling regime we know
how to do this with the help of the renormalization group
equations. As a result, the dependence on $\Lambda_{QCD} \equiv
\Lambda_{PT}$ will finally appear. At the same time, we do not
know how to solve these equations in the strong coupling regime.
So, in order to avoid this problem, we decided to show the
existence of a mass gap explicitly, by extracting the deep IR
asymptotics of the gluon propagator within the separate relevant
diagrams. The rest of the problem is to sum up an infinite number
of the most singular terms (contributions) in order to see whether
or not a mass gap will finally survive.}

\subsection{Gluon confinement criterion}

Precisely this program has been carried out in Refs. \cite{20,21}.
For the sake of completeness, let us repeat it briefly here. We
will show that a mass gap remains, indeed. Thus, on general ground
one has

\begin{equation}
D^{INP}(q^2, \Delta^2) = \sum_{k=0}^{\infty} (q^2)^{-2
-k}(\Delta^2)^{k+1} \sum_{m=0}^{\infty} a_{k,m}(\xi) g^{2m},
\end{equation}
which is the Laurent expansion with respect to the inverse powers
of $q^2$ for the INP part of the full gluon propagator. The
crucial observation is that the regularization of the NP IR
singularities does not depend on their powers \cite{20,21,22},
namely

\begin{equation}
(q^2)^{- 2 - k } = { 1 \over \epsilon} a(k)[\delta^4(q)]^{(k)} +
f.t., \quad \epsilon \rightarrow 0^+,
\end{equation}
where $a(k)$ is a finite constant depending only on $k$ and
$[\delta^4(q)]^{(k)}$ represents the $k$th derivative of the
$\delta$-function. Here $\epsilon$ is the IR regularization
parameter, introduced within the dimensional regularization (DR)
method \cite{23}, and which should go to zero at the end of the
computations. In Ref. \cite{20} it has been proven that neither
$g^2$ nor the gauge fixing parameter $\xi$ is to be IR
renormalized, i.e., they are IR finite from the very beginning.
So, in the Laurent expansion (5.2) the only quantity which should
be IR renormalized is the mass gap itself. It is easy to show that
it is IR renormalized as follows: $\Delta^2 = \epsilon \bar
\Delta^2$ as $\epsilon$ goes to zero. In this case, all the
singularities with respect to $\epsilon$ will be cancelled in the
Laurent expansion (5.2), indeed, and everything will be expressed
in terms of the IR renormalized mass gap $\bar \Delta^2$ only,
which, by definition, exists as $\epsilon$ goes to zero (let us
remind that $g^2 = \bar g^2$ and $\xi = \bar \xi$). Moreover, it
is perfectly clear that only the simplest NP IR singularity will
survive in the $\epsilon \rightarrow 0^+$ limit, namely

\begin{equation}
D^{INP}(q^2, \Delta^2) = \Delta^2 (q^2)^{-2} \sum_{m=0}^{\infty}
a_{0,m}(\xi) g^{2m},
\end{equation}
and all other terms in the expansion (5.2) become terms of the
order of $\epsilon$, at least, in this limit (they start from
$(\Delta^2)^2 \sim \epsilon^2$, while $(q^2)^{-2-k}$ always scales
as $1 / \epsilon$). The so-called "f.t." terms in the
dimensionally regularized Laurent expansion (5.3) after its
substitution into expansion (5.2) become terms of the order of
$\epsilon$, so here and everywhere they vanish in the $\epsilon
\rightarrow 0^+$ limit.

Due to the distribution nature of the simplest NP IR
singularity$(q^2)^{-2}$, which saturates the INP part of the full
gluon propagator, the two different cases should be distinguished.

I. If there is an explicit integration over the gluon momentum
(the so-called virtual gluon due to Mandelstam \cite{19}), then
from Eq. (5.4), on account of Eq. (5.3) with $a(0) = \pi^2$, it
finally follows

\begin{equation}
 D^{INP}(q, \bar \Delta^2)=  \bar \Delta^2 \pi^2
\delta^4(q).
\end{equation}
The $\delta$-type regularization is valid even for the multi-loop
skeleton diagrams, where the number of independent loops is equal
to the number of the gluon propagators. In the multi-loop skeleton
diagrams, where these numbers do not coincide (for example, in the
diagrams containing three or four-gluon proper vertices), the
general regularization (5.3) should be used (i.e., derivatives of
the $\delta$-functions). In Eq. (5.5) an infinite series over $m$
is included into the IR renormalized mass gap $\bar \Delta^2$
(retaining the same notation, for simplicity), making thus it the
UV renormalized and gauge-invariant as well. Let us remind that
some of these quantities are UV divergent.

II. If there is no explicit integration over the gluon momentum
(the so-called actual gluon \cite{19}), then the function
$(q^2)^{-2}$ in Eq. (5.4) cannot be treated as the distribution,
and only the IR regularization of a mass gap comes out into the
play. So the INP part of the full gluon propagator in this case
disappears as $\epsilon $ as $\epsilon \rightarrow 0^+$, namely

\begin{equation}
D^{INP}(q, \bar \Delta^2) \sim \epsilon, \quad \epsilon
\rightarrow 0^+.
\end{equation}
This means that any amplitude for any number of soft-gluon
emissions (no integration over their momenta) will vanish in the
IR limit in our picture. In other words, there are no gluons in
the IR, i.e., at large distances (small momenta) there is no
possibility to observe gluons experimentally as free particles. So
color gluons can never be isolated. This behavior can be treated
as the gluon confinement criterion (see also Ref. \cite{24}).
Evidently, this behavior does not explicitly depend on the gauge
choice in the full gluon propagator, i.e., it is a manifestly
gauge-invariant as it should be. It is worth mentioning that it
coincides with the color confinement criterion proposed in Ref.
\cite{25} (and references therein, see also our paper \cite{20}).

\section{Discussion}

\subsection{A possible generalization}

A tensor structure of each term, which contributes into the NL
part of the full gluon propagator, is absolutely the same as the
tensor structure of the linear contributions, i.e., the quark and
ghost skeleton loops. These structures are given in Eqs. (3.6) and
(3.7). However, the dynamical context of these decompositions,
which is present in the corresponding invariant functions, may be
distinguished. In analogy with the relations (3.6) and (3.7) and
because of the Laurent expansion (5.1), the NL part $T_g[D](q)
\equiv T^g_{\mu\nu}[D](q)$ can be generally decomposed as follows:

\begin{equation}
T^g_{\mu\nu}[D](q) =  \delta_{\mu\nu} \Bigl[ {\Delta^4 \over q^2}
L_g^{(1)}(q^2) + \Delta^2 L_g^{(2)}(q^2) +
 q^2 T_g^{(3)}(q^2) \Bigr]
 +  q_{\mu} q_{\nu} \Bigl[ {\Delta^2
\over q^2} L_g^{(4)} (q^2) + T_g^{(5)} (q^2)\Bigr],
\end{equation}
where $T_g^{(n)}(q^2)$ at $n=3,5$ are invariant dimensionless
functions. They are regular functions of $q^2$, i.e., they can be
present by the corresponding Taylor expansions, but possessing AF
at infinity, and depending thus on $\Lambda_{QCD}$ in this limit.
At the same time, the invariant dimensionless functions
$L_g^{(n)}(q^2)$ at $n=1,2,4$ are to be present by the
corresponding Laurent expansions, namely

\begin{equation}
L_g^{(1,2,4)}(q^2) \equiv L_g^{(1,2,4)}(q^2, \Delta^2) =
\sum_{k=0}^{\infty} (\Delta^2 / q^2)^k a_k^{(1,2,4)},
\end{equation}
where the numbers $a_k^{(1,2,4)}$ by themselves are expansions in
the coupling constant squared (see below). Let us emphasize the
inevitable appearance of the mass gap $\Delta^2$. It characterizes
the nontrivial dynamics in the IR region. This precisely makes the
difference between the linear and NL insertions into the gluon
self-energy. Evidently, this difference is due to different
dynamics: in the linear part there is no explicit direct
interaction between massless gluons, while in the NL part there
is. When the mass gap is zero then this decomposition takes the
standard form. So, the generalization (6.1) makes the explicit
dependence on the mass gap of the full gluon propagator perfectly
clear.

In principle, the estimates like the estimates derived in the
previous Sec. 4 can be extended to the quark and ghost skeleton
loops as well. This means that we can formally generalize the
decompositions (3.6) and (3.7) in a similar way as the
decomposition of the NL part (6.1). However, this is not the case,
since we know the sum of the linear contributions. If the
invariant function $T(q^2)$ in Eq. (3.8) may contain singular with
respect to $q^2$ terms (because of a possible formal
generalization), then the IR singularity of the modified free
gluon propagator (3.3) will be only soften in comparison with the
IR singularity of the free gluon propagator (2.2). In general,
such behavior will only compromise the role of ghosts to cancel
unphysical degrees of freedom of gauge bosons (see explanation
below). That is why we believe that all the dependence on the mass
gap which possibly can come from the quark and ghost skeleton
loops can be factorized into the separate block (the linear part),
which itself cannot produce severe IR singularities in the full
gluon propagator. At the same time, we don't know the sum of the
NL contributions. This also makes the difference between the
linear and NL parts, but now from a mathematical point of view.
Thus, it does not make any sense to generalize the ghost and quark
decompositions (3.6) and (3.7), respectively, in the same way as
the decomposition (6.1). They can contribute into the INP part of
the full gluon propagator numerically only, and as a functions
they contribute into its PT part. Neither ghost nor quark skeleton
loops (3.4) and (3.5), which appear in Eq. (3.1), can cancel its
severely singular behavior in the IR, which was demonstrated
above, and which was due to the pure YM part (i.e., to its NL
part) of the gluon SD equation (3.2).

\subsection{The role of ghosts}

It is well known that in order to maintain the unitarity of
$S$-matrix in QCD the ghosts have to cancel unphysical degrees of
freedom (longitudinal ones) of the gauge bosons \cite{1,7,8}.
Evidently, this is due to the general decomposition of the ghost
skeleton loop (3.6), which shows that it always gives the
contribution of the order $q^2$. In the iteration solution for the
gluon propagator (2.4), $D(q) = D^0(q) - D^0(q)T_{gh}(q)D^0(q)
+D^0(q)T_{gh}(q)D^0(q)T_{gh}(q)D^0(q) + ....$, it cancels one of
$q^2$ in the denominator, which comes from the free gluon
propagator. Thus, each term in this expansion becomes always as
singular as $1/ q^2$. Precisely this makes it possible for ghosts
to cancel, in general, the longitudinal component of the full
gluon propagator, which is, by definition, as singular as $1/
q^2$. From a technical point of view the cancellation can be
explicitly demonstrated in the lowest orders of PT in powers of
the coupling constant squared (see, for example Ref. \cite{7}).
However, this is valid term by term in PT. In other words, in
every order of PT the ghosts will cancel unphysical degrees of
freedom of gauge bosons, making thus them always transverse. The
above-mentioned cancellation in all orders of PT means that it
goes beyond PT. It is a general feature, i.e., it does not depend
on whether the solution, for example to the gluon SD equation is
PT or NP, singular or regular at origin, etc. In other words, the
general role of ghosts should not be spoiled by any truncation
scheme (approach).

On the other hand, the general decomposition (3.6) of the ghost
skeleton loop (3.4) takes place if and only if (iff) the full
ghost propagator (Euclidean signature) $G(k) = - (i / k^2[1 +
b(k^2)])$, where $b(k^2)$ is the ghost self-energy, is as singular
as $1/k^2$ at $k^2 \rightarrow 0$. When the ghost self-energy is
zero, i.e., $b(k^2)=0$., then the full gluon propagator becomes
the free one, i.e., $G(k) \rightarrow G_0(k) = - (i / k^2)$. Thus
the IR singularity of the full ghost propagator cannot be more
severe than the exact IR singularity of the free ghost propagator
in order to maintain the cancellation role of unphysical degrees
of freedom of gauge bosons by ghosts at any nonzero covariant
gauge in all sectors of QCD. There is no way for ghosts to cancel
severe IR singularities, which are of dynamical origin due to the
self-interaction of massless gluons in the true QCD vacuum. There
is no doubt left that the full gluon propagator is essentially
severely modified in the IR because of the response of the NP QCD
vacuum, which is not provided by the PT vacuum.

 However, there
exists one gap in these arguments. If one chooses by hand the
Landau gauge $\xi=0$ from the very beginning, then the unphysical
longitudinal component of the full gluon propagator vanishes. Only
the physical transverse component will contribute to the full
gluon propagator, and it may become regular at zero in this case,
indeed. Otherwise, it is always singular at the origin because the
existence of the longitudinal component always produces, at least,
the IR singularity $1/q^2$ (see Eq. (2.1)). In this case there is
no restriction on the behavior of the ghost propagator in the IR,
and it may become (depending on the truncation scheme) more
singular in the IR than its free counterpart. In Ref. \cite{26}
(and references therein) precisely this type of the solution
(regular gluon propagator and more singular than the free one
ghost propagator) to the system of the SD equations in the Landau
gauge has been found. However, this solution is due to the choice
of the special Landau gauge, so it is a gauge artifact solution.
Being thus a gauge artifact, it can be related to none of the
physical phenomena such as quark and gluon confinement, SBCS, etc,
which are, by definition, manifestly gauge-invariant. At the same
time, gauge artifact solutions may exist as formal solutions to
the SD system of equations. If a regular at zero gluon propagator
will be found in a manifestly gauge-invariant way (i.e., in the
way which does not explicitly depend on the particular covariant
or non-covariant gauge choice), only then it should be taken
seriously into the consideration. To our present knowledge a
manifestly gauge-invariant solution for the smooth gluon
propagator, which will not compromise the general role of ghosts,
is not yet found. Moreover, there exists a serious doubt, in our
opinion, that such kind of the solution can be found at all. Thus,
we are left with singular at the origin gluon propagator, which is
possible in any covariant gauge.

\section{Conclusions}

 Emphasizing the importance of the IR structure of the full gluon
propagator and its close relation to the highly nontrivial
structure of the true QCD ground state, one can conclude:

1). The self-interaction of massless gluons (i.e., the NL
gluodynamics) is responsible for the large scale structure of the
true QCD vacuum.

2). The full gluon propagator in any gauge is inevitably more
singular in the IR than its free counterpart.

3). This requires the existence of a mass gap, which is
responsible for the NP dynamics in the QCD vacuum. It appears on
dynamical ground due to the self-interaction of massless gluons
only. It cannot be interpreted as the effective/dynamical gluon
mass.

 4). Though a mass gap gains contributions from all powers of the
QCD coupling constant squared, it itself plays no any role in the
presence of a mass gap.

 5). We define the NP and the PT IR singularities as more
severe than and as much singular as $1 / q^2$, respectively, which
is the power-type, exact IR singularity of the free gluon
propagator.

6). On this basis we exactly decompose the full gluon propagator
into the INP and PT parts.

7). An IR renormalization of a mass gap only is needed in order to
fix uniquely and exactly the INP part of the full gluon
propagator. It is saturated by the simplest NP IR singularity, the
famous $(q^2)^{-2}$.

8). The main dynamical source of the NP (severe) IR singularities
in the full gluon propagator is the two-loop skeleton term of the
corresponding SD equation, which contains the four-gluon vertices
only.

9). The regular at zero full gluon propagator is to be ruled out.
Only its PT part can be rendered regular at zero due to the
special gauge choice (Landau gauge).

10). As a functions ghost and quark degrees of freedom contribute
only into the PT part of the full gluon propagator within our
approach. As integrated out in all orders of linear PT (i.e.,
numerically) they contribute to its INP part as well.

11). Taking into account the distribution nature of the NP IR
singularities, the gluon confinement criterion is formulated in a
manifestly gauge-invariant way.

 Our general conclusions can be formulated as follows: at the
microscopic, dynamical level just the fundamental NL four-gluon
interaction makes the full gluon propagator so singular in the IR.
This requires the introduction of a mass gap, i.e., it arises
mainly from the quartic gluon potential (see also Refs.
\cite{27,28}).

A financial support from HAS-JINR Scientific Collaboration Fund
(P.Levai) is to be acknowledged.

\end{document}